\documentclass{aa}

\usepackage{epstopdf}
\usepackage{epsfig}
\usepackage{graphics}
\usepackage{natbib}
\usepackage{amsmath}
\usepackage{amssymb} 
\usepackage{enumerate}
\usepackage[toc,page]{appendix}
\begin{document}

\title{Non-thermal emission resulting from a supernova explosion inside an extragalactic jet}

\author{F.~L.~Vieyro\inst{1,2}, V.~Bosch-Ramon\inst{2} \and N.~Torres-Alb\`a\inst{2}}

\institute{Instituto Argentino de Radioastronom\'{\i}a (IAR, CCT La Plata, CONICET; CICPBA), C.C.5, (1984) Villa Elisa, Buenos Aires, Argentina \and Departament de F\'{\i}sica Qu\`antica i Astrof\'{\i}sica, Institut de Ci\`encies del Cosmos (ICC), Universitat de Barcelona (IEEC-UB), Mart\'{i} i Franqu\`es 1, E08028 Barcelona, Spain}

\offprints{F. L. Vieyro \\ \email{fvieyro@iar.unlp.edu.ar}}

\titlerunning{Supernova explosion inside an AGN jet}

%\authorrunning{Vieyro, Bosch-Ramon, Torres-Alb\`a}

\abstract
{Core-collapse supernovae are found in galaxies with ongoing star-formation. In a starburst galaxy hosting an active galactic nucleus with a relativistic jet, supernovae can take place inside the jet. The collision of the supernova ejecta with the jet flow is expected to lead to the formation of an interaction region, in which particles can be accelerated and produce high-energy emission.}
{We study the non-thermal radiation produced by electrons accelerated as a result of a supernova explosion inside the jet of an active galactic nucleus within a star-forming galaxy.}
{We first analyzed the dynamical evolution of the supernova ejecta impacted by the jet. Then, we explored the parameter space using simple prescriptions for the observed gamma-ray lightcurve. Finally, the synchrotron and the inverse Compton spectral energy distributions for two types of sources, a radio galaxy and a powerful blazar, are computed.}
{For a radio galaxy, the interaction between a supernova and a jet of power $\sim 10^{43}-10^{44}$~erg~s$^{-1}$ can produce apparent gamma-ray luminosities of $\sim 10^{42}-10^{43}$~erg~s$^{-1}$, with an event duty cycle of supernova remnant (SNR) interacting with the jet close to one for one galaxy. For a blazar with a powerful jet of $\sim 10^{46}$~erg~s$^{-1}$, the jet-supernova ejecta interaction could produce apparent gamma-ray luminosities of $\sim 10^{43}-10^{44}$~erg~s$^{-1}$, but with a much lower duty cycle.}
{The interaction of supernovae with misaligned jets of moderate power can be relatively frequent, and can result in steady gamma-ray emission potentially detectable for sources in the local universe. For powerful blazars much farther away, the emission would be steady as well, and it might be detectable under very efficient acceleration, but the events would be rather infrequent.}

 
\maketitle

\section{Introduction}

Active galactic nuclei (AGNs) are composed of a supermassive black hole accreting material from the central region of its galaxy host. In order for the black hole to be active, dust, gas, and matter must be available for accretion. One way to enhance accretion is through galaxy mergers \citep{stockton1982}. At the same time, galaxy interactions or mergers can stimulate nuclear star formation \citep{toomre1972}. 

Although galaxies hosting AGNs can be of different types, they tend to be massive galaxies ($M > 10^{10} M_{\odot}$) with younger stellar populations than average \citep{kauffmann2003}. Several studies have shown that luminous AGN hosts are likely to have higher star-formation rates (SFRs) than normal galaxies \citep[e.g.,][]{kauffmann2003,shao2010}; but it may also be the case for low or moderate luminosity AGNs, as suggested by \citet{hickox2014,gurkan2015}. There is growing evidence supporting a close connection between nuclear starbursts and AGNs \citep[e.g.,][]{alexander2012}. For example, AGN surveys show that the fraction of galaxies hosting AGNs is significantly lower for red galaxies (quiescent, composed mainly of old stars) than for blue (star-forming) galaxies \citep{wang2017}. 

Given the large amount of dust released in supernova (SN) explosions, these events were proposed as a possible mechanism to link starburst phenomena and AGN feedback  \citep{ishibashi2016}. Type Ia SNe are the only type of SNe found in old, elliptical galaxies. These are thermonuclear explosions associated either to an accreting white dwarf in a binary system, or to the merger of two white dwarfs \citep[e.g.,][]{maoz2014}. On the other hand, in star-forming galaxies, a high rate of core-collapse SNe resulting from the explosions of massive, short-lived stars is expected \citep{kelly2012}.

Since high SFRs favor the occurrence of  core-collapse SNe, one cannot neglect the possibility of a core-collapse SN taking place inside the jet of a radio-loud AGN. The resulting interaction, with the SN ejecta playing the role of a dynamical obstacle, may lead to detectable non-thermal radiation: When a jet interacts with an obstacle (e.g., clouds, stars with strong winds), a bow-shaped structure forms in the collision region of the two fluids. Particles can be accelerated up to relativistic energies in this region, and produce high-energy emission. The interaction between relativistic jets and obstacles has been explored in several works. Many of these works focused on studying the dynamical effects that the interaction can have on the jet \citep[e.g.,][]{komissarov1994,hubbard06,perucho2014,perucho2017}; others were devoted to study the radiative effects, as gamma-ray flares \citep[e.g.,][]{barkov2010,barkov12b,bosch-ramon2012,banasinski16}, or steady high-energy emission \citep[e.g.,][]{araudo2010,bednarek97,araudo2013,bosch-ramon2015,bednarek2015,wykes15,moreno2016,vieyro2017}. Star-jet interactions were also proposed to explain the rapid-variability observed in some powerful blazars \citep[e.g.,][]{barkov12a,khangulyan13,aharonian2017}.

In the present work, we have considered the dynamics and radiation of an AGN jet interacting with a SN. At first, the explosion is not halted by the jet presence until the SN ejecta becomes diluted enough. Then, when the SN ejecta reaches a size similar to the stagnation radius, that is, when the SN ejecta and jet ram pressures balance each other, the SN ejecta expansion is significantly slowed down by the jet impact in all directions but downstream the jet. From that point on, the evolution of the supernova remnant is strongly affected by the jet impact. The possibility of supernova-AGN jet interaction has been previously explored for instance by \cite{blandford79}, who considered the prospect of such an interaction being the cause of the knots observed in the jet of the galaxy M87. In addition, \cite{bednarek1999} discussed the possibility of very efficient particle acceleration due to the interaction of an AGN jet with a SN shell. In that work, the SN was assumed to be within or close enough to the jet for the interaction to be significant, although the dynamical evolution of the interaction and the associated gamma-ray emission were not computed.

This article is organized as follows: In Section~\ref{dynamic}, the dynamical evolution of a SN ejecta accelerated by a jet is described using a simple model; this model is complemented in the Appendix with hydrodynamical simulations. In Sect.~\ref{appL}, we explore the jet parameter space, and study the outcomes of different scenarios using a simple prescription for the gamma-ray lightcurve. In Sect.~\ref{SED} we compute the spectral energy distributions (SEDs) resulting from the electrons accelerated in the jet-SN ejecta interaction; we explore two cases: a nearby radio galaxy, and a more distant and powerful blazar. We discuss the adopted model and its results in Sect.~\ref{discussion}, and close with the conclusions in Sect.~\ref{conclusions}.

\section{Dynamical evolution}\label{dynamic}

The dynamical evolution of an obstacle, for example, a cloud of gas, inside a jet has been extensively studied in several works. The impact of the jet causes a transfer of momentum and the consequent acceleration of the cloud. In addition, it produces a shock wave that propagates trough the cloud, compressing and heating it up. The heated  material suffers a lateral expansion occurring approximately at the speed of sound. The cloud also
forms an elongated tail in the downstream direction as a result of the pressure gradient  caused by the impact of the jet. Kelvin-Helmholtz instabilities will start striping material from the cloud downstream. Rayleigh-Taylor instabilities can also develop at the cloud surface directly impacted by the jet, given the acceleration exerted by the jet on the cloud. Eventually, the instabilities should lead to the disruption of the cloud, mixing the latter with the jet flow. 

The above description is rather simplified; in reality it is far more complex. Numerical studies show that for quite homogeneous clouds, the cross section can grow significantly before the cloud total disruption \citep{bosch-ramon2012}. This takes longer than the time it takes the shock to cross the cloud \citep[e.g.,][]{cooper2009,pittard2010}, and the (initial) acceleration timescale of the cloud \citep{bosch-ramon2012}. Radiative cooling, accompanied by subsequent cloud disruption, has been shown to significantly extend in time the obstacle role of a cloud impacted by a supersonic, non-relativistic wind \citep{cooper2009}. On the other hand, relativistic simulations still show efficient expansion and acceleration of the cloud despite its disruption and radiative cooling \citep{perucho2017}. The evolution can also be altered by additional factors, such as magnetic fields or thermal conduction, that are not taken into account in this work \citep[e.g.,][]{klein94,fragile05}.

For the acceleration of the cloud, we have based our study on \citet{barkov12b}. In that work, the authors describe  a model for the acceleration of a cloud, in the present case the SN ejecta, pushed by a magnetically-dominated jet. Here, we studied the dynamical evolution in the case of a purely hydrodynamical jet, since on the jet scales of interest most of the jet magnetic energy is expected to have been already transferred to kinetic energy.

We refer to the laboratory and the SN ejecta reference frame  $K$ and $K'$ , respectively. For a relativistic jet with negligible thermal pressure, the momentum flux of the jet in $K'$ is:
\begin{equation}
f' = \Gamma_{\rm rel}^2 \beta_{\rm rel}^2 \rho_{\rm j} h_{\rm j} c^2 ,
\end{equation}
where $h_{\rm j} = 1+ \hat{\gamma} \epsilon_{\rm j}$ is the jet enthalpy, $\hat{\gamma}$ the adiabatic index (4/3 and 5/3 for a relativistic and a non-relativistic ideal, monoatomic adiabatic gas, respectively), and $\beta_{\rm rel} = (\beta_{\rm j} - \beta_{\rm c})/(1- \beta_{\rm j} \beta_{\rm c})$ the relative velocity between the jet and SN ejecta in $c$ units, which have velocities $\beta_{\rm j}$ and $\beta_{\rm c}$ in $K$, respectively. The jet momentum flux in $K$, $f$, relates to $f'$ through:
\begin{equation}\label{eq:qqp}
f' = \Big(1 - \frac{\beta_{\rm c}}{\beta_{\rm j}} \Big)^2 \Gamma_{\rm c}^2 f . % \beta_{\rm rel}^2 \frac{1}{\Gamma_{\rm j}^2-1} q .
\end{equation}
\noindent where $\Gamma_{\rm c}$ is the Lorentz factor of the cloud.

The SN ejecta momentum increases due to the acceleration caused by the jet in its direction of motion. The equation of motion is \citep{barkov12b}:
\begin{equation}
\frac{d \Gamma_{\rm c}}{dt} = \frac{\pi r_{\rm c}^2 \beta_{\rm c}}{M_{\rm c}c} f',
\end{equation}
which, combined with Eq.~(\ref{eq:qqp}), results in:
\begin{equation}\label{eq:G_evol_hydro}
\frac{d \Gamma_{\rm c}}{dt} = 
\frac{\pi r_{\rm c}^2 \beta_{\rm c}}{M_{\rm c}c}\Big(1 - \frac{\beta_{\rm c}}{\beta_{\rm j}} \Big)^2 \Gamma_{\rm c}^2 f
\approx\frac{L_{\rm j}}{M_{\rm c} c^2}\Big(\frac{r_{\rm c}}{r_{\rm j}(z)}\Big)^2 \Big( 1- \frac{\beta_{\rm c}}{\beta_{\rm j}} \Big)^2 \Gamma_{\rm c}^2 \beta_{\rm c}\,,
\end{equation}
where $L_{\rm j}=\pi r_{\rm j}^2 \Gamma_{\rm j}^2 \beta_{\rm j} \rho_{\rm j} h_{\rm j} c^2$ ($\approx \pi r_{\rm j}^2f\,c$ for $\Gamma_{\rm j}\gg 1$) is the jet power (including rest mass). 

The jet is assumed to be conical, that is, with a constant opening angle $\theta$, where the radius of the jet is a function of the distance $z$ to the black hole: $r_{\rm{j}}(z)\approx \theta\,z$. We adopt a jet Lorentz factor of $\Gamma_{\rm j} =10$ and an opening angle of $\theta = 1/ \Gamma_{\rm j}$.

The time lasted by the event as seen by the observer relates to the time in the laboratory frame $t$ as 
 \begin{equation}\label{eq:tobs}
t_{\rm obs}(z) =\int_{z_0}^z (1-\beta_{\rm c} \cos i ) dt=\int_{z_0}^z (1-\beta_{\rm c} \cos i ) \frac{dz}{\beta_{\rm c}c}\,,
\end{equation}
where $i$ is the angle between the jet axis and the line of sight.  

The impact of the jet, in addition to transferring kinetic energy, heats up the SN ejecta as well. The SN ejecta is in pressure balance with the shocked jet flow that is pushing, with the pressure at the contact region being $\sim f'$. Since initially the jet flow is strongly supersonic in $K'$, $f'$ is larger than the jet lateral total pressure, and the jet-SN ejecta pressure balance leads to the above mentioned fast SN ejecta expansion, fueled by the jet-transferred heat. 

The expansion of the SN ejecta enhances the interaction with the jet, favoring the acceleration of the former, and also its disruption. As already noted, simulations show that the (deformed) SN ejecta expansion and acceleration can continue for some time (\citealt{bosch-ramon2012}; \citealt{perucho2017}; see also the Appendix). If the SN ejecta achieves a relativistic regime, which is the case for a powerful jet, its expansion in the laboratory frame is slowed down: (i)  in the jet direction, by the relatively small velocity difference between different parts of the SN ejecta; (ii) in the direction perpendicular to the jet, relativistic time dilation in the flow frame leads to a slow expansion. We note that the lateral pressure of the shocked jet fluid, which passes around the SN ejecta, may contribute to confine the SN ejecta, slowing its expansion down to some extent (not considered here but at the end of the expansion; see Sect.~\ref{apdymo}). These effects combined allow the SN ejecta to keep some integrity even when close to total disruption, extending the time needed for fully mixing with the jet flow and the traveled distance inside the jet in the laboratory frame. For weak jets, the SN ejecta accelerates at a low rate, and it is expected to cover a longer distance inside the jet before its disruption \citep{khangulyan13}. These predictions are supported by results presented in the Appendix, where we show the results of an axisymmetric, relativistic hydrodynamical simulation of an interaction between the SN ejecta and a relatively weak jet. Based on these result, and on the above discussion for the powerful jet case, we have assumed for simplicity here that most of the SN ejecta mass remains in causal contact with the jet contact surface, effectively evolving as a roughly spherical cloud, with its radius increasing as
\begin{equation}
r_{\rm c}(t) = R_0 + \int \frac{c_{\rm s} dt}{\Gamma_{\rm c}(t)},
\end{equation}
where $R_0$ is the initial SN ejecta radius, and 
\begin{equation}
c_{\rm s}^2 = \frac{ \hat{\gamma} P_{\rm c} }{h_{\rm c} \rho_{\rm c} }
\end{equation}
is the SN ejecta sound speed squared, with $h_{\rm c}$ being the specific enthalpy of the SN ejecta, and $P_{\rm c}$ its pressure. In the present context, $R_0$ is determined through balancing its pressure and $f'$. This condition can be written as
\begin{equation}\label{eq:ro}
\left[P_{\rm c}=f'\right]\,\sim\,\left[\frac{3E_0}{10 \pi R_0^3} = \frac{L_{\rm j}}{c \pi r_{\rm j}(z)^2}\right],
\end{equation}
where $E_0$ is the total energy of the SN ejecta, adopted as the standard isotropic SN luminosity of $E_0 = 10^{51}$~erg. For simplicity, we adopted a reference SN ejecta mass value of $M_{\rm c} = 10\,M_{\odot}$. We note that, since  $\beta_{\rm c} \sim 0$ at the beginning of the interaction, we use $f' = f$ in Eq.~(\ref{eq:ro}).
 
\subsection{Applicability of the dynamical model}\label{apdymo}

When the SN ejecta has expanded significantly, its pressure can become smaller than the jet lateral pressure, taken here one hundredth of $f$ as a fiducial value. Its exact value does not have a strong impact on the results as long as the jet is highly supersonic in $K$. Once the SN ejecta pressure is equal to the jet lateral pressure, the SN ejecta evolves more smoothly with the jet flow. From that point on, we assumed that the jet energy and momentum transferred to the SN ejecta become small, and therefore neglect any further acceleration of particles.

When the SN ejecta covers the whole jet section, its expansion rate is assumed to be the same as that of the jet, i.e., $r_{\rm c}\sim r_{\rm j}$, with the jet flow at the interaction location moving with the SN ejecta. This is a reasonable assumption to zeroth order, as long as the jet external medium is much denser than the jet itself, which is expected for a jet on the galactic disk scales, as the denser medium inertia encapsulates the jet shocked flow (see Sect.~\ref{discussion}). An accurate account of this situation can show complex features in the jet and SN ejecta hydrodynamics; this requires a  numerical study, and a detailed account of such a process is out of the scope of this work (see the Appendix for a simulation with a simple gas model). 

\section{General study}\label{appL}

\subsection{Simplified model}\label{appL_rad}

The galaxy host is assumed to be a starburst with a high star-formation rate ($\dot{M}_{\rm SFR}$) and a disk geometry. We considered that the starburst has an IR luminosity of $L_{\rm IR} =10^{12} L_{\odot}$ \citep[e.g., an ultraluminous infrared galaxy, ULIRG,][]{sanders1988}. The stellar and IR fields are modeled as gray bodies with characteristic temperatures of $T_{\rm s} = 30000$~K and $T_{\rm IR} = 200$~K, respectively \citep{vieyro2017}.

There are three important free parameters in the model, $L_{\rm j}$, $\Gamma_{\rm j}$, and $i$. Throughout this work, we consider the jet Lorentz factor to be constant; in particular, we adopt $\Gamma_{\rm j}=10$, which is a common value in AGNs \citep[e.g.,][]{jorstad2017}. Moderate variations around this value do not affect significantly the results of the model for jets with modest power, but the effect would be important for powerful blazars (see Sect.~\ref{vargam}). The parameter space of the jet luminosity and inclination can be studied adopting a simplified model of the apparent non-thermal luminosity. The limitations of this simplified model are discussed in Sect.~\ref{discussion}.

We considered that the SN explosion takes place at a height $z_0$ inside the jet. We fix $z_0\sim 50$~pc, as these are roughly the scales at which it is more likely that a SN explosion will occur within the jet, for the adopted star-forming disk. For smaller $z_0$, the jet volume is much smaller; for larger $z_0$, in the disk periphery and beyond, star formation is suppressed. 

Inside the jet, the shocked SN ejecta acceleration and evolution results in an evolution of the jet energy flux dissipated at the jet-SN ejecta interaction surface in $K'$. This dissipated energy flux of the jet can be taken as a proxy of the energy flux injected into non-thermal particles. The corresponding power injected into non-thermal particles at $z$, due to the jet-SN ejecta interaction, can be expressed in $K'$ as:
\begin{equation} 
\label{eq:appLnt}
L'_{\rm NT}(z) = \eta_{\rm NT}\,\pi\,r_{\rm c}^2\,\beta_{\rm rel}\Gamma_{\rm rel}\,(\Gamma_{\rm rel}\,h_{\rm j}-1)\, \rho_{\rm j}\,c^3,
\end{equation}
where the constant $\eta_{\rm NT}$ is the fraction of jet energy that impacts the SN ejecta and is converted into non-thermal energy. We note that the rest-mass energy has been removed to derive $L'_{\rm NT}$. A reference value of 0.1 has been adopted for $\eta_{\rm NT}$ in this work; high enough so significant radiation is predicted, but well below the upper-limit of 1 (see Sect.~\ref{impetant}).

Only a fraction of the energy that is injected in non-thermal particles is channeled into radiation. This fraction is quantified with the radiative efficiency $\xi'_{\rm IC/sync}(E',z)$, which in  $K'$ can be estimated as
\begin{equation}\label{eq:frad}
\xi'_{\rm IC/sync}(E',z) = \frac{t'^{-1}_{\rm IC/sync}}{t'^{-1}_{\rm rad} + t'^{-1}_{\rm nrad}}\,,
\end{equation}
where $t'^{-1}_{\rm nrad}$ accounts for the non-radiative losses in $K'$ (e.g., adiabatic losses and particle advection, of similar scale; see \citealt{vieyro2017}), and  $t'^{-1}_{\rm rad}=t'^{-1}_{\rm IC}+t'^{-1}_{\rm sync}$ for synchrotron and IC losses in $K'$. The emitted energy in $K'$ is, then, $L'_{\rm intr} = L'_{\rm NT}(z)\xi'_{\rm IC/sync}(E',z)$.

The intrinsic luminosity should be corrected by Doppler boosting. The IC and synchrotron lightcurves of the radiation as seen by the observer can be estimated as
\begin{equation}
\label{eq:Ldoppler}
L_{\rm IC/sync}^{\rm app}(t_{\rm obs}) = \delta_{\rm c}(z)^4 L'_{\rm intr} = \delta_{\rm c}(z)^4L'_{\rm NT}(z) \xi'_{\rm IC/sync}(E,z)\,,
\end{equation}
where $\delta_{\rm c}$ is the Doppler boosting factor of the emitting flow:
\begin{equation}\label{eq:doppler}
\delta_{\rm c} = \frac{1}{ \Gamma_{\rm c}(1-\beta_{\rm c} \cos i )}\,,
\end{equation}
and $t_{\rm obs}$ is related to z through Eq.~(\ref{eq:tobs}). Equation~(\ref{eq:Ldoppler}) is valid as long as the accelerated particles follow an energy distribution similar to $\propto E^{-2}$, meaning that the energy is equally distributed among different energy scales (see also Sect.~\ref{SED}).

We have focussed here on electrons (and positrons) as radiating particles, and synchrotron and IC as radiative processes, as they emit the most efficiently in the regions of interest. Previous works considered also hadronic emission close to or at the jet base \citep[e.g.,][]{barkov2010,barkov12a}.

We derived the magnetic field to compute synchrotron emission assuming that the total magnetic energy density is a fraction $\zeta_{\rm eq}$ of the jet energy density. For a magnetic field predominantly perpendicular to the flow motion (e.g., toroidal), in $K'$ one obtains:
\begin{equation}\label{eq:Bfield}
B'_{\phi}(z) \approx \frac{1}{\Gamma_{\rm c} z}\sqrt{\frac{4\zeta_{\rm eq}L_{\rm{j}}}{\theta^2c}}\,.
\end{equation}
An equipartition magnetic field, $\zeta_{\rm eq}=1$, would imply in our convention that the jet energy density is equally divided between magnetic and particle energy density. Throughout the article we adopt $\zeta_{\rm eq}=10^{-2}$ (except when otherwise indicated), which results in a magnetic field 10 times below the equipartition value, as estimated in some extragalactic jets \citep{hardcastle2011}. 

We consider as main target photons for IC interactions the infrared (IR) field associated with starburst galaxies \citep{vieyro2017}, and compute the IC cooling rate using the full Klein-Nishina cross section \citep{khangulyan2014} and following the target treatment described in \citet{dermer1993}.

To obtain the lightcurves of IC emission, we calculate $L_{\rm IC}^{\rm app}$ at a reference energy of $E'_{\rm IC} = (m_e c^2)^2/kT_{\rm IR}\Gamma_{\rm c}$, where $E'_{\rm IC}$ is approximately at the maximum of the IC cross section in $K'$ for (quasi)head-on IC interactions, around the transition from the Thomson to the Klein-Nishina regime.

\subsection{Results of the simplified model}\label{appL_res}

\begin{figure}[!htb]
\centering
%\hfill
\includegraphics[width=0.45\textwidth,keepaspectratio]{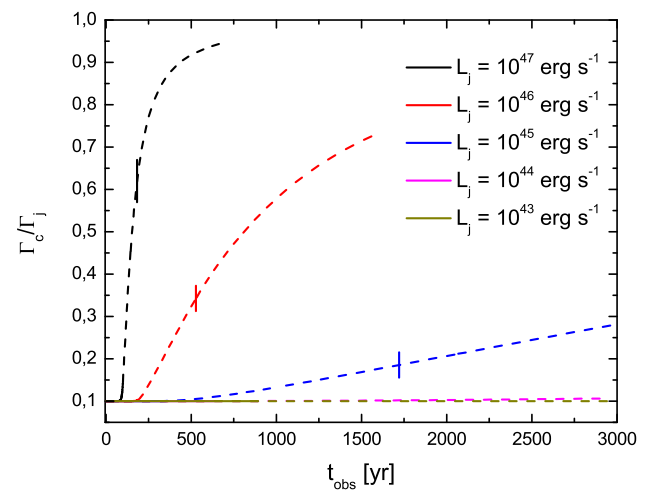}\hspace{20pt} 
\caption{Evolution of the SN ejecta Lorentz factor for different jet powers, for  $i=0^{\circ}$. The vertical ticks mark the point where the SN ejecta crosses 1~kpc. The lines become dashed when the SN ejecta covers the whole jet section.}
\label{fig:Gc}
\end{figure}

In Fig. \ref{fig:Gc}, we show the evolution of the SN ejecta Lorentz factor for blazar-like sources (i.e., $i=0^{\circ}$), for different values of jet power, $\Gamma_{\rm j} =10$ and $z_0 = 50$~pc. The more powerful the jet, the shorter the time it takes for the SN ejecta to be accelerated to a higher Lorentz factor; it reaches a relativistic regime before covering the whole jet section only for the most powerful jets ($L_{\rm j} = 10^{46}-10^{47}$~erg~s$^{-1}$). We recall that, for all the cases studied, $\eta_{\rm NT}=0.1$.

In all cases, the SN ejecta expands and eventually covers the whole jet cross section. After this point, the dynamical model becomes less suitable to describe the evolution of the SN ejecta, hence in the figure we show this evolution phase using dashed lines (see Sect.~\ref{discussion} for a discussion of the validity of the model). 

\begin{figure}[!ht]
\centering
%\hfill
\includegraphics[width=0.45\textwidth,keepaspectratio]{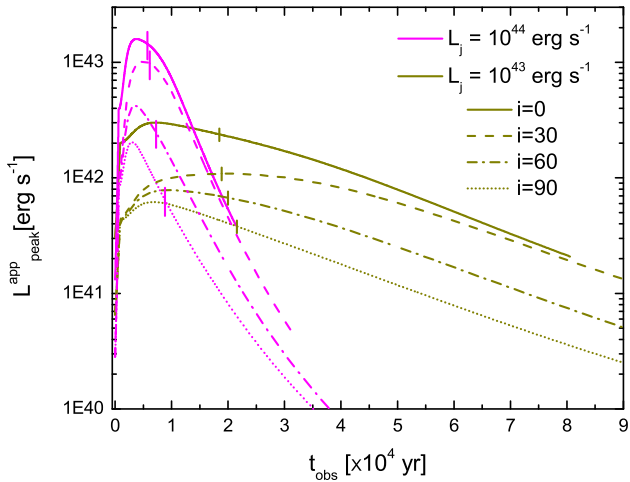} \hspace{20pt} 
\caption{Gamma-ray IC lightcurves computed for different inclination angles. As in Fig. \ref{fig:Gc}, the vertical bars indicate the point where the SN reaches 1~kpc.}
\label{fig:25kpc}
\end{figure}

In Fig.~\ref{fig:25kpc}, we show the approximated IC lightcurve in gamma rays, as seen by the observer, for $L_{\rm j} = 10^{43}-10^{44}$~erg~s$^{-1}$ and different inclination angles. For intermediate inclinations, say $i=30^{\circ}$, the apparent luminosities are above $10^{42}$ and $10^{41}$~erg~s$^{-1}$, during an observed period of $\sim 10^4$ and $\sim 10^5$~yr, for $L_{\rm j} = 10^{44}$ and $10^{43}$~erg~s$^{-1}$, respectively. For completeness, the lightcurves show the SN ejecta propagating until it reaches $z=25$~kpc. Nevertheless, the jet properties can change significantly on kpc scales (e.g., the jets may be already disrupted in weak, FRI-type jets), and our prescription for the SN ejecta evolution may be far from correct in those regions. Therefore, effective jet-SN ejecta interaction may be reliable up to $z \lesssim 10$~kpc, and in fact results beyond $z\sim 1$~kpc (indicated in the figures by vertical ticks) should be taken with caution.

\begin{figure}[!t]
\centering
%\hfill
\includegraphics[width=0.45\textwidth,keepaspectratio]{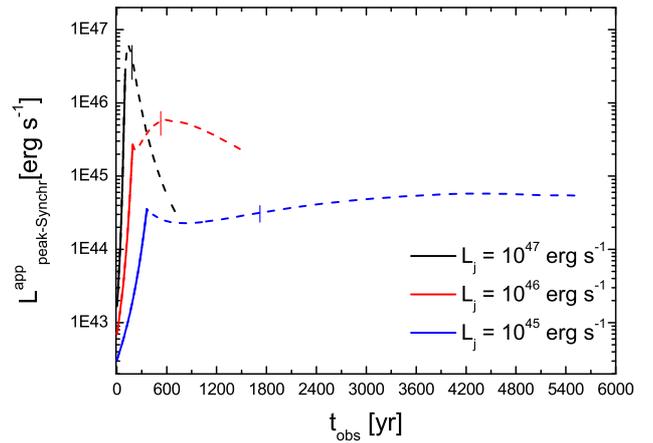} \hspace{20pt} 
\caption{Gamma-ray synchrotron lightcurves for powerful jets with $i=0^{\circ}$. As in Fig. \ref{fig:Gc}, the solid-to-dashed transition indicates the moment the SN ejecta covers the jet section, and the vertical bars show the point where the SN crosses 1~kpc.}
\label{fig:Lsyn}
\end{figure}

For the most powerful jets, $L_{\rm j}  \geq 10^{45}$~erg~s$^{-1}$, synchrotron self-Compton (SSC) is the main mechanism of gamma-ray emission and for such process the semi-analytical approach described in Sect.~\ref{appL_rad} is no longer valid. For these jets, however, the approach is still valid for, and worth applying to, the synchrotron gamma-ray emission, which may reach $\gtrsim 100$~MeV as seen by the observer. The critical synchrotron energy can be estimated as
\begin{equation}
E_{\rm c} = \varsigma  \textrm{ } 236 \textrm{ MeV } \delta_{\rm c}, 
\end{equation}
where $\varsigma<1$, determines the particle acceleration rate $\dot{E}=\varsigma\,q\,B\,c$, and typically is not well constrained. For values of $\varsigma \gtrsim 1/\delta_{\rm c}$, the critical energy can reach values of $\sim 100$~MeV as seen by the observer. Figure~\ref{fig:Lsyn} shows the synchrotron lightcurves for the powerful jets; we consider blazar-like sources, since high $\delta_{\rm c}$ values are necessary to obtain photons of $100$~MeV as seen by the observer. As in Fig. \ref{fig:25kpc}, the evolution is computed until $z=25$~kpc, and the vertical bars mark the moment when the jet crosses 1~kpc. It can be seen that for these jets, the synchrotron radiation could dominate the gamma-ray emission in the {\it Fermi} energy range. As mentioned in Sect.~\ref{appL_rad}, we have adopted $\zeta_{\rm eq}=10^{-2}$; since synchrotron losses are dominant for high energy electrons, the luminosities obtained at $\sim 100$ MeV are similar for $\zeta_{\rm eq}=10^{-4}-1$.

\begin{figure}[!t]
\centering
%\hfill
\includegraphics[width=0.45\textwidth,keepaspectratio]{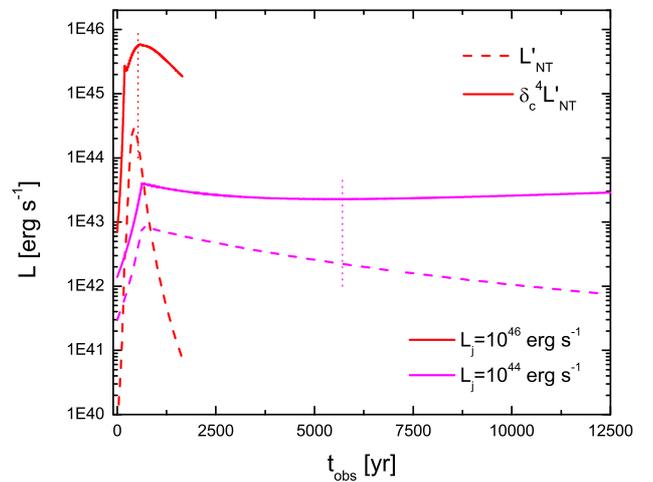} \hspace{20pt} 
\caption{Luminosity injected into non-thermal particles in $K'$, $L'_{\rm NT}$ (dashed lines), and corrected by Doppler boosting, $L'_{\rm NT} \delta_{\rm c}^4$ (solid lines), for $L_{\rm j} = 10^{44}-10^{46}$ erg s$^{-1}$, and $i=0^{\circ}$. The vertical bars show the point where the SN ejecta crosses 1~kpc.}
\label{fig:Lapp_Lnt}
\end{figure}

In Fig. \ref{fig:Lapp_Lnt} we compare the power injected into non-thermal particles in the fluid frame, $L'_{\rm NT}$, and corrected by Doppler boosting, $L'_{\rm NT} \delta_{\rm c}^4$, for $L_{\rm j} = 10^{44}$ erg s$^{-1}$ (IC lightcurve) and $L_{\rm j} = 10^{46}$ erg s$^{-1}$ (synchrotron lightcurve). In both cases, the rapid rise in the luminosity at the beginning of the event is caused by the increase of the flux injected into non-thermal particles (given by Eq. \ref{eq:appLnt}). The SN ejecta reaches low Lorentz factors for the weakest jets (at least within 1~kpc, as can be seen in Fig. \ref{fig:Gc}), thus Doppler boosting is not strong, and it only moderately increases the luminosity, as seen by the observer. For the strongest jets ($\gtrsim 10^{45}$ erg s$^{-1}$), Doppler boosting is more relevant throughout the interaction. The higher peak luminosities and slower decay in $L'_{\rm NT} \delta_{\rm c}^4$ are the result of Doppler boosting in all the cases.

In \citet{barkov12b}, the characteristic effective timescale of the interaction, that is the time during which the SN ejecta-jet interaction intensity is strong enough for effective particle acceleration to occur, can be roughly estimated as
\begin{equation}\label{eq:dt}
\Delta t \sim 10^3 \Big( \frac{\Gamma_{\rm j}}{10} \Big) \Big( \frac{L_{\rm j}}{10^{46} \textrm{ erg s}^{-1}} \Big)^{-1} \Big( \frac{M_{\rm c}}{10M_{\odot}} \Big) \textrm{ yr}\,.
\end{equation}
This expression is valid once the SN ejecta covers the complete jet section, which occurs early in all the cases studied here. This is in agreement with the results shown in Fig.~\ref{fig:Lsyn} for powerful jets. For weaker jets, the slow acceleration of the SN ejecta renders Eq.~(\ref{eq:dt}) not suitable to estimate $\Delta t$.

\section{Spectral energy distributions}\label{SED}

As illustrative cases, the synchrotron and IC SEDs are computed at the time when the SN ejecta covers the jet section for two cases: a nearby radio galaxy with an intermediate jet power, and a powerful blazar at higher redshift. Both galaxies are considered to host a nuclear starburst as described in Sect.~\ref{appL}. We considered the radio galaxy to be located in the local universe at a distance $d = 100$~Mpc ($z\approx 0.026$, for $H_0=70$~km~s$^{-1}$~Mpc$^{-1}$, $\Omega_{\Lambda} =0.7$, and $\Omega_{\rm M} =0.3$), with a jet power $L_{\rm j} = 10^{44}$~erg~s$^{-1}$, and an inclination angle  $i=60^{\circ}$. On the other hand, the more powerful blazar is considered to be located at $z=1$ (equivalent to a luminosity distance of $6.6$~Gpc), hosting a powerful jet with $L_{\rm j} = 10^{47}$~erg~s$^{-1}$ pointing toward the observer. Table~\ref{table1} lists all the relevant parameter values of the model and the sources.

\begin{table}[ht]
    \caption[]{Main parameters of the model.}
        \label{table1}
        \centering
\begin{tabular}{lcc}

\hline\hline  \\ [0.005cm]

Parameters & Radio galaxy  & Blazar \\ [0.005cm]

\hline\\ [0.005cm]

$L$: jet power [erg s$^{-1}$]    & $10^{44}$    & $10^{47}$     \\
$\Gamma_{\rm j}$: jet Lorentz factor    & $10$          & $10$  \\
$i$: inclination angle  & $60$  & $0$            \\
$z$: redshift   & $0.026$       & $1$   \\  %$0.012$

\hline \hline \\[0.005cm]

\end{tabular}
\end{table}

In order to compute the SEDs, we assume an injection rate of non-thermal particles in $K'$ following an energy distribution of the form:
\begin{equation}
Q'(E',z) = Q_0(z) E'^{-\alpha} \exp(-E'/E'_{\rm max}(z))\,,
\end{equation}
where $\alpha=2$ is taken as a fidutial value, typical for efficient accelerators, and characteristic of diffusive acceleration mechanisms. Functions $Q$ much softer in energy would lead to gamma-ray emission much more difficult to detect, whereas harder $Q$ would slightly increase the gamma-ray output. The value of $E'_{\rm max}$, determining the maximum particle energy, has been derived as in \cite{vieyro2017}. The total non-thermal luminosity injected is: 
\begin{equation}
\int Q'(E',z)E'dE'=L_{\rm NT}'(z)\,. 
\end{equation}

\begin{figure*}[!ht]
\centering
%\hfill
\includegraphics[width=0.45\textwidth,keepaspectratio]{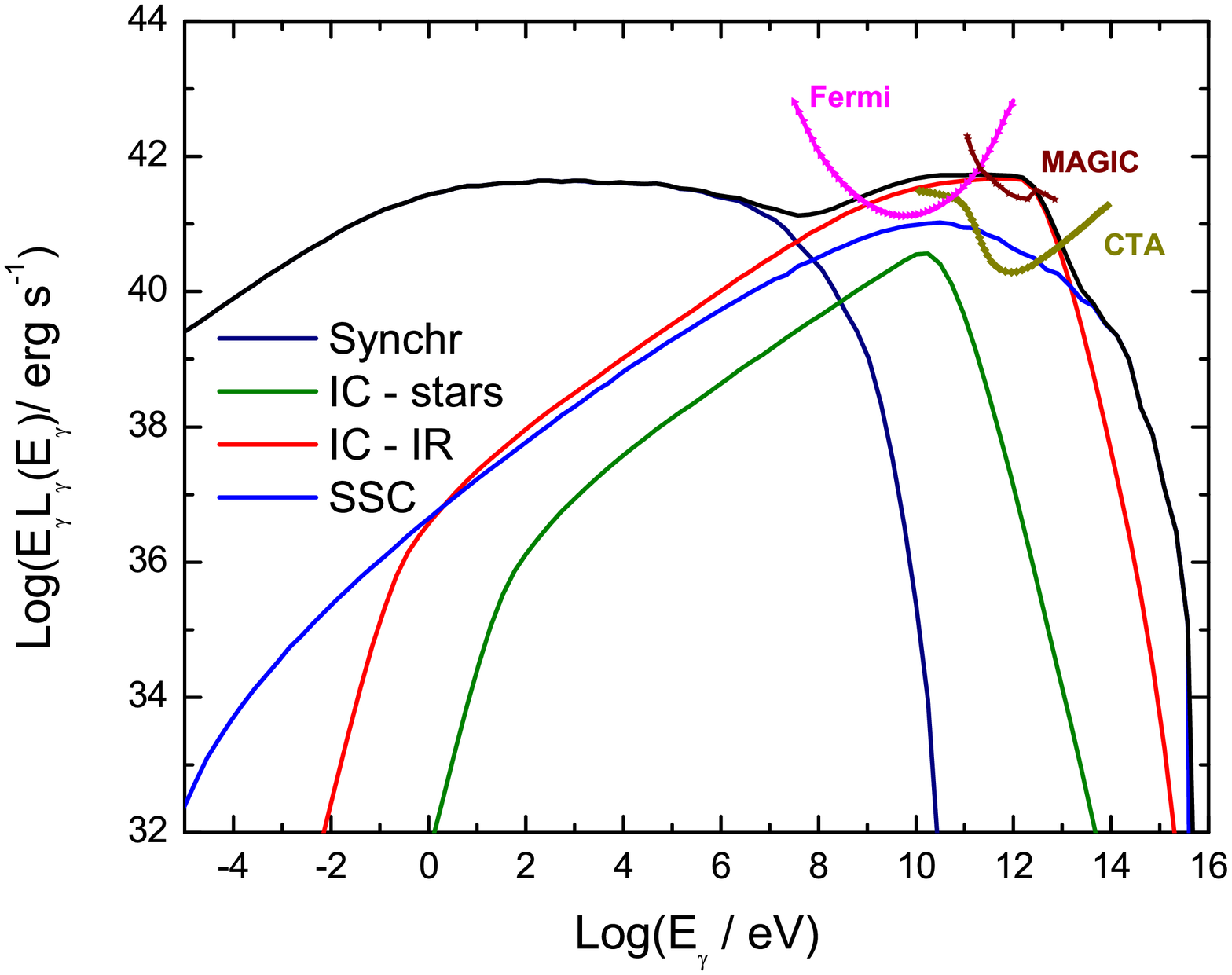} \hspace{20pt} 
\includegraphics[width=0.45\textwidth,keepaspectratio]{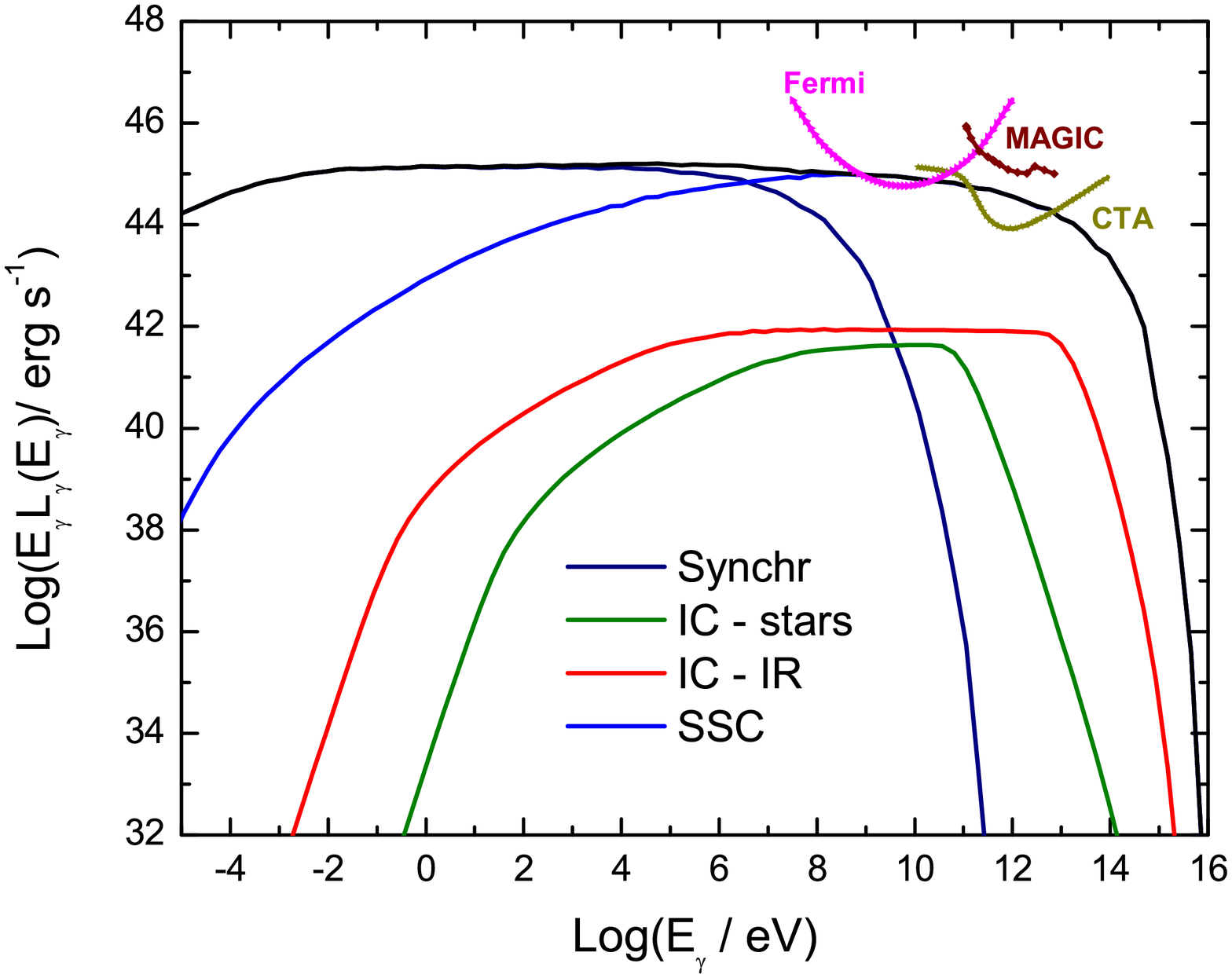}\hspace{20pt} 
\caption{Synchrotron and IC SEDs for the radio galaxy (left) and the blazar (right) cases, together with the sensitivities of different gamma-ray instruments ({\it Fermi} -pink-, presently operating Cherenkov telescopes -brown-, CTA -green-), at $z=51.3$ and $55.3$~pc, respectively.}
\label{fig:radio_vs_blazar}
\end{figure*}

For each $z$ value, the transport equation in steady state is solved for an homogeneous emitter (one-zone), which has the following semi-analytical solution:
\begin{equation}
N'(E',z) = \frac{1}{ |\dot{E}'(E',z)|}  \int_{E'}^{E'_{\rm max}}  {Q'(E^*,z) dE^*}\,,
\end{equation}
where $|\dot{E}'(E',z)| = E'\,t'^{-1}_{\rm nrad+rad}(E',z)$ accounts for the radiative and the non-radiative electron energy losses \citep{vieyro2017}. We consider three different target fields for IC interactions: the radiation from the stars in the galaxy, IR photons associated with the starburst, and synchrotron emission (for SSC). The SSC calculations are correct as long as IC losses of synchrotron targets are not dominant. Although for powerful jets SSC is the main mechanism for gamma-ray emission (as shown in Fig.~\ref{fig:radio_vs_blazar}), it is not the dominant radiative loss mechanism.

Figure~\ref{fig:radio_vs_blazar} shows the computed SEDs obtained for the radio galaxy (left panel) and the blazar (right panel), together with the sensitivities of the {\it Fermi} observatory, MAGIC, as an example of current imaging air Cherenkov telescopes (IACTs), and the future Cherenkov Telescope Array (CTA). These SEDs correspond to the moment when the SN ejecta covers the whole jet section (see the solid-to-dashed line transition in Fig.~\ref{fig:Gc}). For the radio galaxy this occurs at $z=51.3$~pc, when the SN ejecta has a Lorentz factor of only $\Gamma_{\rm c} = 1.0003$, whereas for the blazar, the Lorentz factor is $\Gamma_{\rm c} = 1.43$, at  $z=55.3$ pc. Although the peak in the lightcurves is predicted to be somewhat later in both cases (as seen, for example, in Fig.~\ref{fig:Lsyn} for the blazar case), we compute the SEDs when $r_{\rm c}$ equals  $r_{\rm j}$ because, up to this point, the semi-analytical treatment for the SN ejecta evolution is reasonably accurate. Nevertheless, the difference in the luminosity levels in the lightcurves between the moment when $r_{\rm c}=r_{\rm j}$, and their maxima, is only a factor of approximately two to three. Gamma-ray absorption effects are important above 10~TeV and have not computed for the SEDs, but they are discussed in Sect.~\ref{grabs} below.
 
\subsection{Main observational characteristics}

The radio galaxy case yields the most optimistic predictions for detection, as its IC emission may be detectable by {\it Fermi} and current IACTs, for the adopted parameter values. In the blazar case, the synchrotron losses are dominant, causing a significant decrease in the gamma-ray IC luminosity. This effect, together with the large distances involved (say $z\sim 1$), makes the jet-SN ejecta interactions difficult to detect from very powerful blazars with current instruments, but potentially detectable by CTA in the future. We note that a blazar of intermediate power, say $L_{\rm j}\sim 10^{44}$~erg~s$^{-1}$, would resent a similar SED to the one of the radio galaxy (Fig.~\ref{fig:radio_vs_blazar}, left panel), with a higher normalization due to beaming effects. This case could be easily detected in the local universe, although this kind of source is rare with respect to sources of equal power with misaligned jets. 

Very bright(or weak) radio sources may be evidence of the presence of high (/low) magnetic fields, and can be used to constrain the $\zeta_{\rm eq}$ parameter. Low X-ray fluxes may also be an indicator of a very low magnetic fields. These comparison between bands may be difficult for weak radio and X-ray emission as it could be easily masked by other persistently emitting regions. 

We can compare our predictions on different wavelengths with the steady emission detected from sources with similar characteristics to those of the two examples studied here. For instance, the well known quasar 3C~273 has a jet close to the line of sight with an estimated power of $L_{\rm j} \sim 10^{46}$~erg~s$^{-1}$. The host galaxy is classified as an ULIRG, implying high SFR and IR luminosity. The gamma-ray luminosity observed by {\it Fermi} during a quiescent state in 2009 \citep{abdo2010} is $L_{1-10 \rm{GeV}} \gtrsim 10^{45}$~erg~s$^{-1}$, comparable to the blazar case (right panel of Fig~\ref{fig:radio_vs_blazar}). For the adopted value of $\zeta_{\rm eq}=10^{-2}$, the radio fluxes obtained are also similar to the typical observed fluxes from 3C~273, whereas in X-rays intrinsic jet emission, or even an accretion disk (as the one in 3C~273), could hide the radiation from a jet-SN ejecta interaction. In the radio galaxy case we take M87 as reference. This galaxy has a jet of $L_{\rm j} \sim 10^{44}$~erg~s$^{-1}$, with an inclination angle of $i=20^{\circ}$. The detected radio luminosity is of $L_{230 \rm{GHz}} \sim 7 \times 10^{40}$~erg~s$^{-1}$ \citep{DoeFis2012}, similar to the one obtained here. In the $\sim$GeV range we also obtained fluxes comparable to those observed in the steady state of M87. In X-rays, however, our predictions are greater than the fluxes of M87; this could be alleviated by reducing $\zeta_{\rm eq}$, which would affect also the predicted radio luminosity, or adopting a much lower value for $\varsigma$ (the acceleration rate efficiency). 

The observer luminosities predicted in this work for jet-SN ejecta interactions are comparable to those already observed in steady sources, and particular spectral shapes cannot be predicted from a purely phenomenological particle acceleration model (one may say that typical acceleration spectra render typical radiation spectra). The magnetic field strength is also difficult to assess from first principles, adding more freedom to the spectral outcome of synchrotron and IC. In addition, the long timescales involved imply that the predicted lightcurves cannot be distinguished from intrinsic jet persistent emission. We can state, however, that SN ejecta are arguably the largest effective internal obstacles that AGN jets can encounter. Anything as massive, such as a compact molecular cloud, will be too diluted to fully enter the jet, whereas smaller objects such as stars and their winds can hardly cover a whole jet but in rare occasions: a large wind momentum rate plus a weak jet; lighter obstacles will produce also shorter events. Finally, for relatively nearby sources, radio VLBI could be used to resolve the obstacle, and discriminate different scenarios. For instance,  \cite{muller14} found evidence of jet-obstacle interaction (probably with a star) in Centaurus~A (see also \citealt{snios2019}). A detailed case-by-case study, rich in observational information (not very common), is needed to ascertain whether a particular source persistent activity may be associated to a jet-SN ejecta interaction.

\subsection{Gamma-ray absorption}\label{grabs}

\begin{figure*}[!ht]
\centering
%\hfill
\includegraphics[width=0.45\textwidth,trim={0cm 5cm -1.5cm 2cm},angle=270,keepaspectratio]{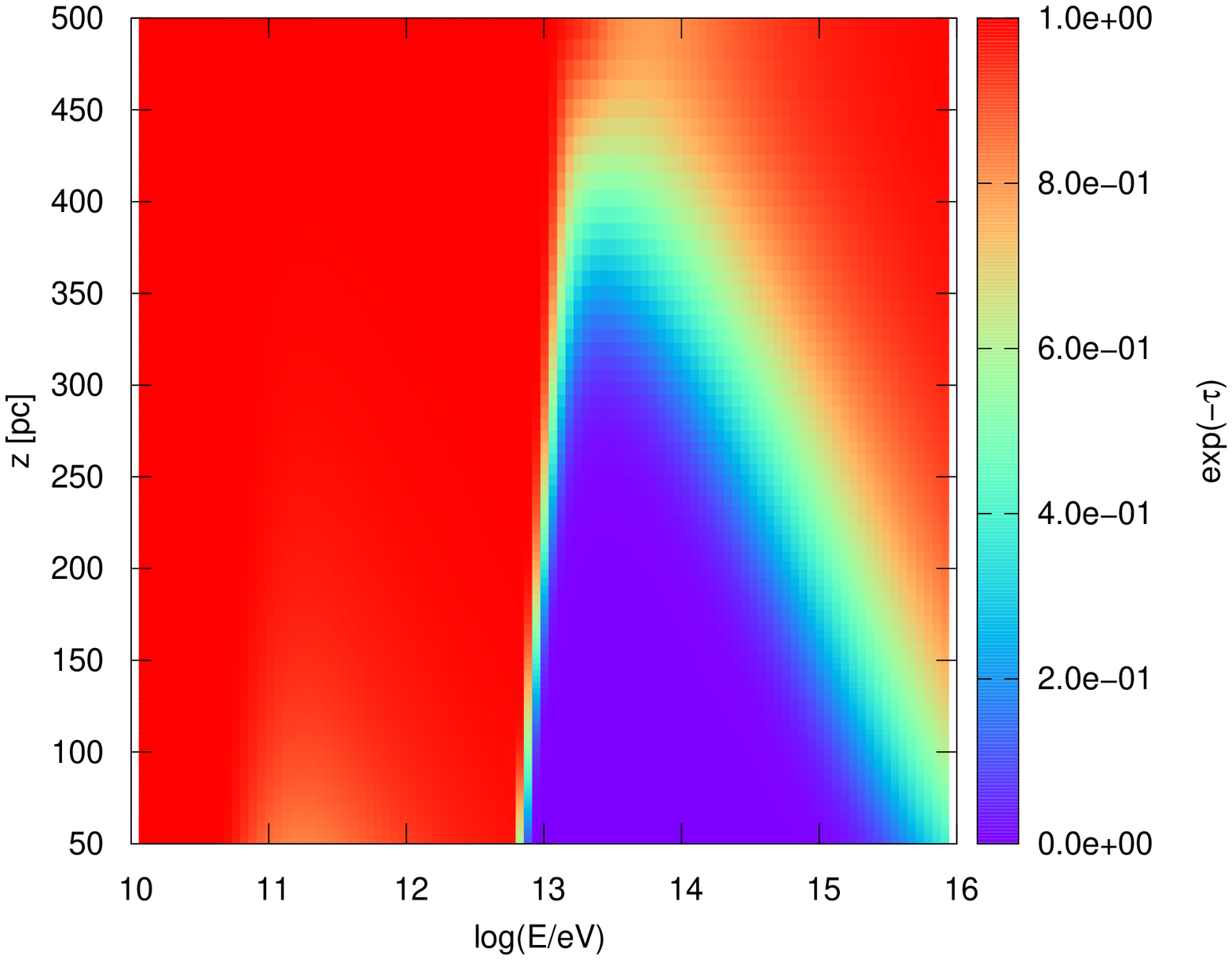} \hspace{20pt} 
\includegraphics[width=0.45\textwidth,trim={0cm 5cm -1.5cm 2cm},angle=270,keepaspectratio]{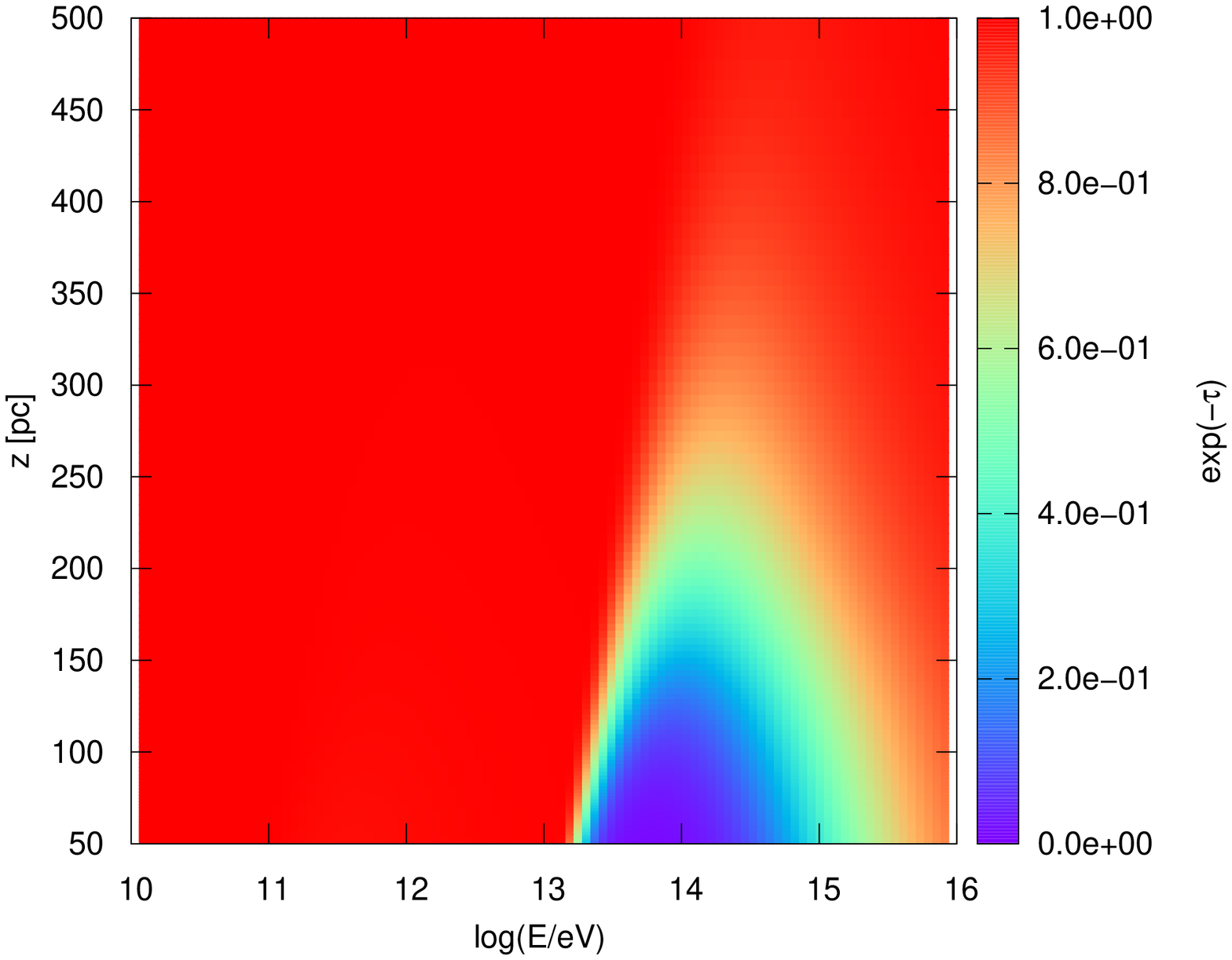}\hspace{20pt} 
\caption{Absorption maps in the IR and OB star radiation fields, for the radio galaxy (left panel) and the blazar (right panel).}
\label{fig:abs}
\end{figure*}

Gamma-ray absorption cannot be neglected in the explored scenario in the IR and UV fields of the starburst. Figure~\ref{fig:abs} shows maps of gamma-ray opacity, associated to $e^\pm$ pair creation in the IR and OB star radiation fields, for the radio galaxy (left panel) and the blazar (right panel) cases. Absorption above $\sim 10$~TeV is due to the IR field of the starburst. OB star emission affects mostly gamma rays of energies $\gtrsim 100$~GeV.

The IR absorption can be roughly estimated by $\tau_{\gamma \gamma} \sim 0.2 \sigma_{\rm T} n_{\rm IR} R_{\rm d}$, where $n_{\rm IR} = L_{\rm IR}/\pi R_{\rm d}^2 c (2.7 k T_{\rm IR})$. For the values adopted in Sect.~\ref{appL_rad}, we obtain $\tau_{\gamma \gamma} \sim 80$. This implies that all the emission at energies $\gtrsim 10$~TeV should be absorbed when the SN is within or close to the starburst disk, which is the case for the considered events around their maxima. This absorption would lead to pair creation in the jet surroundings, with a subsequent secondary synchrotron and IR-target IC emission. Given the complex structure of the jet-SN ejecta interaction region, it is difficult to assess the anisotropy level of the secondary radiation, as some pairs may get boosted if injected in the unshocked jet, while others would get isotropized in the surroundings. We speculate that, in the blazar scenario, this secondary emission may be minor with respect to the overall beamed emission, whereas for radio galaxies this contribution may be more important (see, e.g., \citealt{inoue2011}).

The UV field of OB stars, unlike the IR field, has a considerably lower impact. The optical depth takes now values of $\lesssim 10^{-2}$ for the radio galaxy, and even lower for the blazar case. This absorption should lead to secondary pairs emitting at $\sim 10$ GeV energies, although their contribution would probably be minor.

\section{Discussion}\label{discussion}

In this work, we studied the interaction of a relativistic jet with a SNR and its radiative consequences. First, we  estimated in a simplified manner the observed gamma-ray luminosity evolution expected from this interaction. We then calculated, with more detail, the SED expected for a radio galaxy at $d=100$~Mpc and a blazar at $z=1$.  We discuss below some of the assumptions adopted in this work.

\subsection{Model comparison}\label{comp}

\begin{figure}[!ht]
\centering
%\hfill
\includegraphics[width=0.4\textwidth,keepaspectratio]{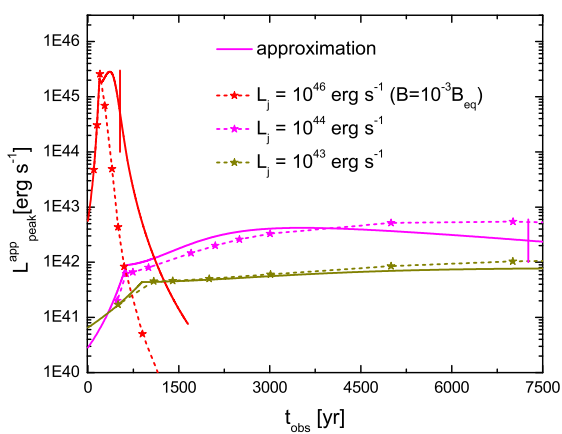} \hspace{20pt} 
\caption{Comparison between the gamma-ray lightcurves obtained with the simplified prescription (solid curve) and the more detailed treatment (dashed curve; stars indicate the points where the luminosity was computed). We consider $i=60^{\circ}$ and $\zeta_{\rm eq}=10^{-2}$ for $L_{\rm j} = 10^{43}$ and $10^{44}$~erg~s$^{-1}$, and $i=0^{\circ}$ and $\zeta_{\rm eq}=10^{-6}$  for $L_{\rm j} = 10^{46}$~erg~s$^{-1}$. The vertical ticks indicate the moment the SN reaches 1~kpc (for $L_{\rm j} = 10^{43}$~erg~s$^{-1}$ this takes place at $t_{\rm obs} \sim 1.8 \times 10^4$~yr, hence it is not shown in the plot).}
\label{fig:Lapp_SED}
\end{figure}

In Fig.~\ref{fig:Lapp_SED} we compare the luminosity evolution obtained using the simplified treatment given by Eq.~(\ref{eq:Ldoppler}) and the luminosity computed as described in Sect.~\ref{SED}. The plot corresponds to jets with $L_{\rm j} = 10^{43}-10^{44}$~erg~s$^{-1}$, $i=60^{\circ}$ and $\zeta_{\rm eq}=10^{-2}$. We have also included a comparison for a more powerful jet, of $L_{\rm j} = 10^{46}$~erg~s$^{-1}$ and $i=0^{\circ}$, for which we considered a well below equipartition magnetic field, $B=10^{-3}B_{\rm eq}$, in order for the external IC to dominate the emission over SSC. For the less powerful, non-blazar jets, the simple analysis predicts the emission rather accurately; for the powerful blazar, the simple prescription overestimates the luminosity by approximately an order of magnitude (as found already in \citealt{vieyro2017}). The discrepancy of the synchrotron emission around $\sim 100$~MeV predicted for blazar-sources by the two approaches (not shown in the figure) is higher than for the IC emission; during the peak of the event, however, we obtain the same difference of $\sim 0.1$.

\subsection{Nature of the emitting flow}

The assumptions that the emitter moves with the SN ejecta and has its size are in fact  assumptions whose validity depends on the scenario. When the SN ejecta is slow, it efficiently acquires momentum, but not energy, while for a faster and diluted SN ejecta the energy transfer becomes more efficient \citep[see][for a discussion of the energy transfer phases]{barkov2010}. Therefore, for a slow SN ejecta with a radius smaller than that of the jet, the shocked jet flow can be far more efficient at dissipating jet power in the form of non-thermal particles than within the SN ejecta. The emission from this quasi-stationary flow will be beamed, which would not be the case for an emitter moving with the SN ejecta. On the other hand, in the subrelativistic regime, the emitter is expected to be larger than the SN ejecta due to the extended oblique shock farther downstream \citep{bosch-ramon2015}. Thus, the predicted radiation luminosity is affected by these unaccounted factors: beaming from a quasi-stationary flow, and a larger dissipation region.  

When the SN ejecta becomes relativistic, the energy transferred from the jet to the shocked jet flow and to the SN ejecta becomes similar, and thus the latter may become a significant emitter in addition to the shocked jet flow. Both the shocked jet flow and the SN ejecta will still have different Doppler boosting patterns until $\Gamma_{\rm c}\rightarrow \Gamma_{\rm j}$, point at which particle acceleration should become very weak or null. 

Despite the qualitative differences, in the context of phenomenological modeling the model and most of the parameter values are basically the same regardless of the actual emitting flow: the shocked jet flow or the SN ejecta. The magnetic field values may otherwise differ in both regions.

In the scenario we studied, the radius of the SN ejecta tends to the jet radius. If the SN ejecta covers the whole jet section well before $\Gamma_{\rm c}\rightarrow \Gamma_{\rm j}$, the shocked jet flow is still the most significant emitter, but it will move at the same velocity as the SN ejecta as long as the shocked jet flow is encapsulated by the external medium around the SN ejecta (see the following discussion and the Appendix).

\subsection{A jet covered by the SN ejecta}

As mentioned in Sect.~\ref{dynamic} the shape of the SN ejecta impacted by the jet grows sidewards and forms an elongated tail along the jet direction. The ability of the SN ejecta to intercept jet energy is the most important factor in our study, and this depends exclusively on the size of the SN ejecta perpendicular to the jet. The lateral expansion of the SN ejecta near the surface impacted by the jet is expected to be fast. Here, for simplicity, we assumed that this expansion takes place at the sound speed, and that the density evolves as if the SN ejecta were spherical. SN ejecta disruption, plus some lateral pressure exerted by the shocked jet, makes this approximation less accurate. However, as discussed in the Appendix, the evolution of the main parameters obtained with the semi-analytical approach does not deviate significantly from the results of axisymmetric, relativistic hydrodynamical simulations.\footnote{We note here the complexities and uncertainties related to implementing an accurate numerical approach, which makes a semi-analytical treatment valuable for exploring several cases.}.

Expansion leads the SN ejecta to cover the whole jet section. At that point, we assume that the SN ejecta is confined by the jet walls and expands at the same rate as the conical jet. For a jet propagating in vacuum this approximation would be wrong, since nothing would prevent the SN ejecta to expand further. However, extragalactic jets usually propagate in media much denser than the jet itself, in particular when crossing their host galaxies. Thus, the dense external medium, heated and compressed by the SN ejecta when $r_{\rm c}\gtrsim r_{\rm j}$, strongly slows down its expansion with its large inertia. At most, the speed of the lateral expansion of the SN ejecta should be that of the Sedov-Taylor phase, with $\dot{R}_{\rm c,st}\sim (L_{\rm j}/\rho_{\rm ISM})^{1/5}t^{-2/5}$. For most cases, after just $\sim 100$~yr, $\dot{R}_{\rm c,st}$ already becomes lower than $\theta\beta_{\rm j}c$, meaning that the SN ejecta and the jet do eventually expand at roughly the same rate. Such a situation is likely to prevent the SN ejecta from expanding sidewards beyond the jet. If $\Gamma_{\rm c}\rightarrow\Gamma_{\rm j}$, this situation may not have such a strong impact on the jet global structure, although the medium-SN ejecta interaction could slow down the latter. For slow or slowed-down SN ejecta, the braked jet should become disrupted at the $z$ of the interaction with the SN ejecta, and shocked jet material flowing backwards may strongly affect jet propagation even far upstream, filling a lobe-like structure. The SN ejecta should also get disrupted by the jet impact, although at a slower rate, at least when the SN ejecta density is still higher than the jet density (e.g., \citealt{bosch-ramon2012}; see also the Appendix). 

Some of the effects discussed can be found in the complementary simulation presented in the Appendix. An accurate treatment of this scenario, combining detailed hydrodynamical simulations and precise radiation calculations, is left for future work.

\subsection{Varying $\Gamma_{\rm j}$ and $M_{\rm c}$}\label{vargam}

The jet Lorentz factor adopted in this work, $\Gamma_{\rm j}=10$, was taken as a reference value for illustrative purposes, but adopting $\Gamma_{\rm j}=5$ does not have a significant impact on the results for the weakest jets. We note that the velocity of the jet is likely non uniform through the section of the latter. The jet is expected to develop a shear layer as a transition region to the external medium. In addition, it has been proposed that the jet could consist of a light, ultra-relativistic, electron-positron pair plasma central spine, and a hadronic, heavier and slower outer layer, resulting from the Blandford-Znajek and Blandford-Payne processes (e.g., \citealt{xie2012,ghisellini2005} and references therein). A non-uniform velocity profile with jet radius may not have a major impact in the explored scenario: the SN ejecta completely covers the jet before getting relativistic speeds, and thus the effect of any radial profile of the jet properties should tend to get smoothed out over the jet-SN ejecta contact surface. In any case, a non-homogeneous jet thrust and energy flux is likely to affect the SN ejecta evolution (e.g., enhancing instability growth), which is worth of a devoted future numerical study.

We also considered the impact of a less massive SN ejecta, adopting $M_{\rm c}=1 M_{\odot}$. The main difference obtained in this case is the duration of the events, which become approximately ten times shorter than for $M_{\rm c} = 10\,M_{\odot}$, for the same $L_{\rm j}$-value; the lightcurve peak luminosities are, on the other hand, similar.

\subsection{Impact of $\eta_{\rm NT}$}\label{impetant}

An additional free parameter of the model is the acceleration efficiency $\eta_{\rm NT}$. Here we have adopted a constant value of $0.1$ throughout the paper. This parameter at present can only be constraint observationally, with a range as wide as $\eta_{\rm NT}=0-1$, and its value may also change as the properties of the jet-SN ejecta contact region evolve. Here, our results simply scale linearly with $\eta_{\rm NT}$, and any change in the efficiency linearly affects the predicted luminosities. We note that additional acceleration sites may be present as well, as for instance: the more oblique region of the jet-SN ejecta shock present when $r_{\rm c}\ll r_{\rm j}$ and $\Gamma_{\rm c}\ll \Gamma_{\rm j}$; or the region encompassing the SN-ejecta, the jet termination, and the external medium when $r_{\rm c}\gtrsim r_{\rm j}$. Here, we have considered only the jet-SN ejecta interaction region with section $\sim \pi\,r_{\rm c}^2$.

\subsection{Duty cycle}\label{dc}

To determine how frequent the jet-SN ejecta interactions are in an AGN hosting a central disk-like starburst, one can estimate the SN rate expected in this type of galaxy. Stars with initial masses $M>8 M_{\odot}$ end their life as core-collapse SN \citep{matzner1999}; the upper limit on the progenitor mass is not clearly determined, but there is evidence that massive stars, with $M \gtrsim 20 M_{\odot}$, collapse into a black hole, failing in produce a SN \citep{smartt2015}. We considered an initial mass function $\phi(m) \propto m^{-\alpha}$, where $\alpha=0.3$ for $0.01 \leq m/M_{\odot} < 0.08$, $\alpha=1.8$ for $0.1 \leq m/M_{\odot} < 0.5$, $\alpha=2.7$ for $0.5 \leq m/M_{\odot} < 1$, and $\alpha=2.3$  for $m/M_{\odot} \geq 1$ \citep{kroupa2001}. Assuming a constant SFR, the core-collapse SN rate can be estimated according to \citep{mattila2001}:
\begin{equation}\label{eq:rate}
R_{\rm SN} = \dot{M}_{\rm SFR} \frac{\int_{8 M_{\odot}}^{20 M_{\odot}} \phi(m) dm }{\int_{0.1 M_{\odot}}^{120 M_{\odot}} m \phi (m) dm }.
\end{equation}
For $\dot{M}_{\rm SFR} = 100 M_{\odot}$ yr$^{-1}$, the SN rate in the starburst disk results in $\sim 70$ SN per century.

For a starburst disk with radius $R_{\rm d}=300$~pc and total thickness $h_{\rm d}=100$ pc, only $\sim 0.01$\% of these SNe will take place inside the jet with the adopted geometry. As discussed in Sect.~\ref{appL_res}, for a jet power $L_{\rm j}\sim 10^{43}-10^{44}$~erg~s$^{-1}$, a non-blazar source (say $i\sim 30^\circ$), and $\eta_{\rm NT}=0.1$, the interaction could result in a gamma-ray luminosity $\gtrsim 10^{41}-10^{42}$~erg~s$^{-1}$ for periods of $\sim 10^4$ yr. This implies that for a single radio galaxy, the duty cycle of core-collapse SNe exploding within the jet should be approximately one.
These gamma-ray luminosities may be detectable by {\it Fermi} and current IACTs, and in the future by CTA, for sources up to a few hundreds of Mpc, perhaps even further away for a more extreme choice of parameter values (e.g., $\eta_{\rm NT}\rightarrow 1$). In addition, provided the high duty cycle, several of these sources in the sky may be simultaneously producing gamma rays due to jet-SN ejecta interactions. 

Regarding blazar type sources, the most powerful ones, say $L_{\rm j}\sim 10^{47}$~erg~s$^{-1}$, may be detectable at $z\sim 1$ in the future by CTA. However, the brief nature of such events, with a lightcurve peak duration $\sim 100$~yr and a duty cycle per source of $\sim 1$\%, and the scarcity of objects, would imply a low frequency of occurrence. 

As future work, we plan to study the statistics of starburst AGN with jet-SN ejecta interactions. In addition, Type Ia SNe can also occur in non star-forming galaxies, which are much more numerous than galaxies hosting starbursts. The interaction between a jet of an AGN and a Type Ia SN should also be studied, as many AGN are massive elliptical galaxies with jets. This study is also work under way.

\section{Conclusions}\label{conclusions}

In galaxies with high SFRs and jets of moderate power, the duty cycle of the interaction of the jet with SNe could be close to unity. This implies a rather steady gamma-ray luminosity that may be detectable, perhaps by {\it Fermi} and current IACTs, and more likely by the future instrument CTA, for sources in the local universe. Since there are several nearby galaxies with the characteristics assumed in this work, jet-SN ejecta emission could be responsible for some of the radio galaxies and relatively weak blazars detected as persistent gamma-ray sources. Blazars with powerful jets, not common in the local universe, might be still detectable at farther distances due to the expected higher luminosities, although the shorter duration of the events and scarce object numbers make their detection more unlikely.

\section*{Acknowledgments}
We want to thank an anonymous referee for his/her useful comments that helped to improve the manuscript. This work was supported by the Spanish Ministerio de Econom\'{i}a y Competitividad (MINECO/FEDER, UE) under grants AYA2013-47447-C3-1-P and AYA2016-76012-C3-1-P with partial support by the European Regional Development Fund (ERDF/FEDER), MDM-2014-0369 of ICCUB (Unidad de Excelencia `Mar\'{i}a de Maeztu'), and the Catalan DEC grant 2014 SGR 86. F.L.V acknowledges support from the Argentine Agency CONICET (PIP 2014-00338). N.T.A. acknowledges support from MINECO through FPU14/04887 grant.

\bibliographystyle{aa}
\bibliography{myrefs9}   %expects file references.bib

\begin{appendix}\label{appendix1}

\section{Two-dimensional hydrodynamical simulations}

\begin{figure*}[!ht]
\centering
%\hfill
\includegraphics[width=0.4\textwidth,keepaspectratio]{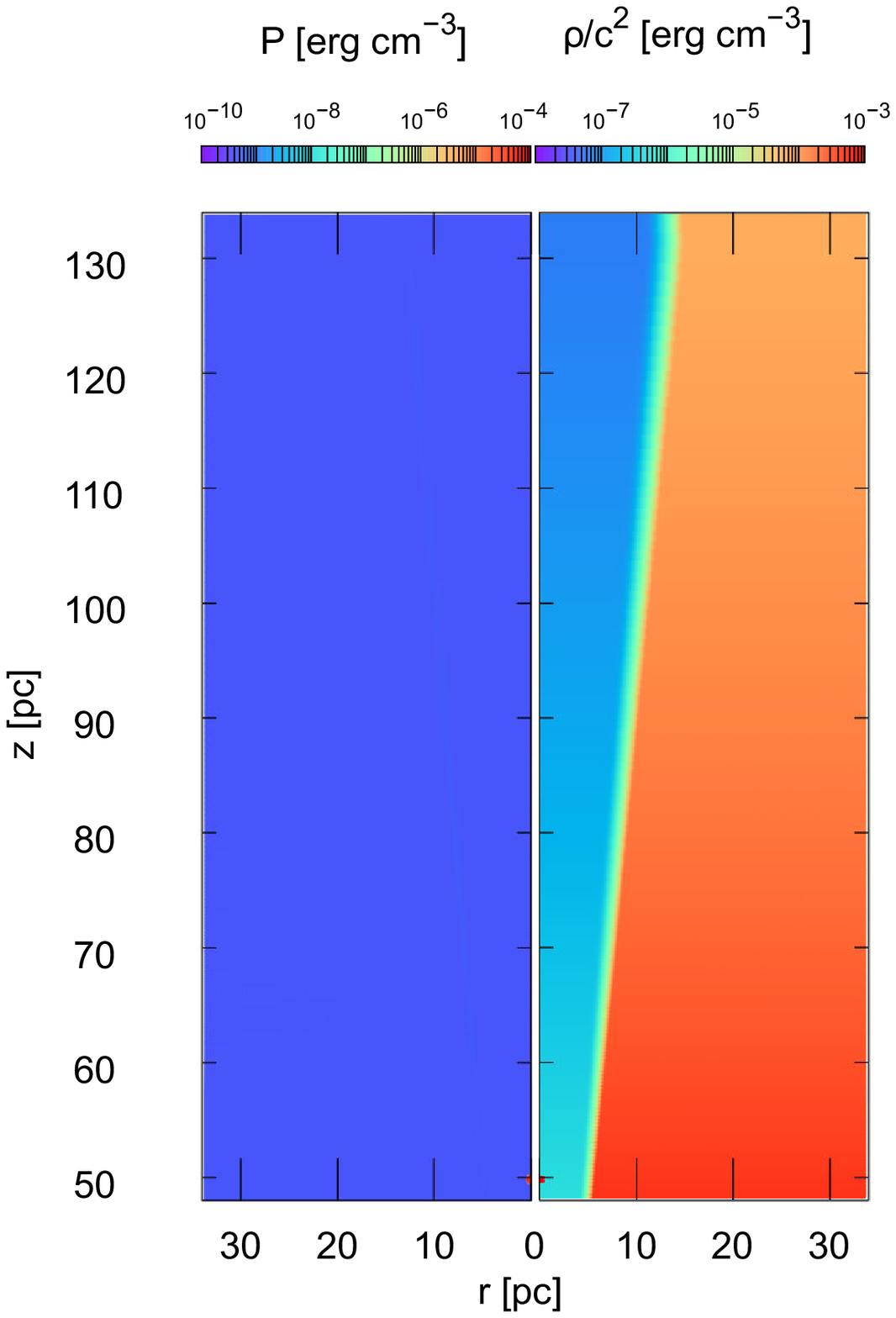} \hspace{20pt} 
\includegraphics[width=0.4\textwidth,keepaspectratio]{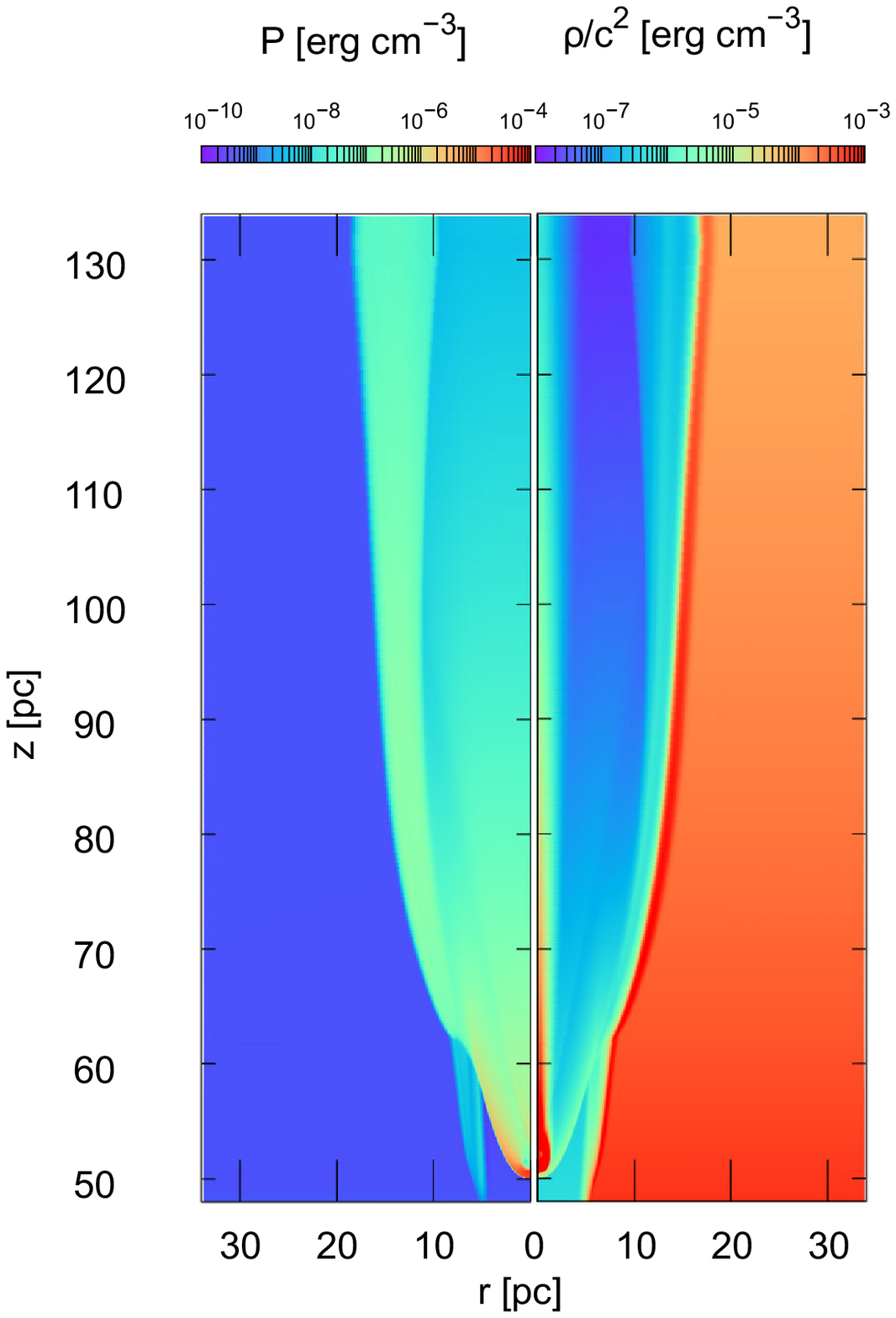} \hspace{20pt} 
\includegraphics[width=0.4\textwidth,keepaspectratio]{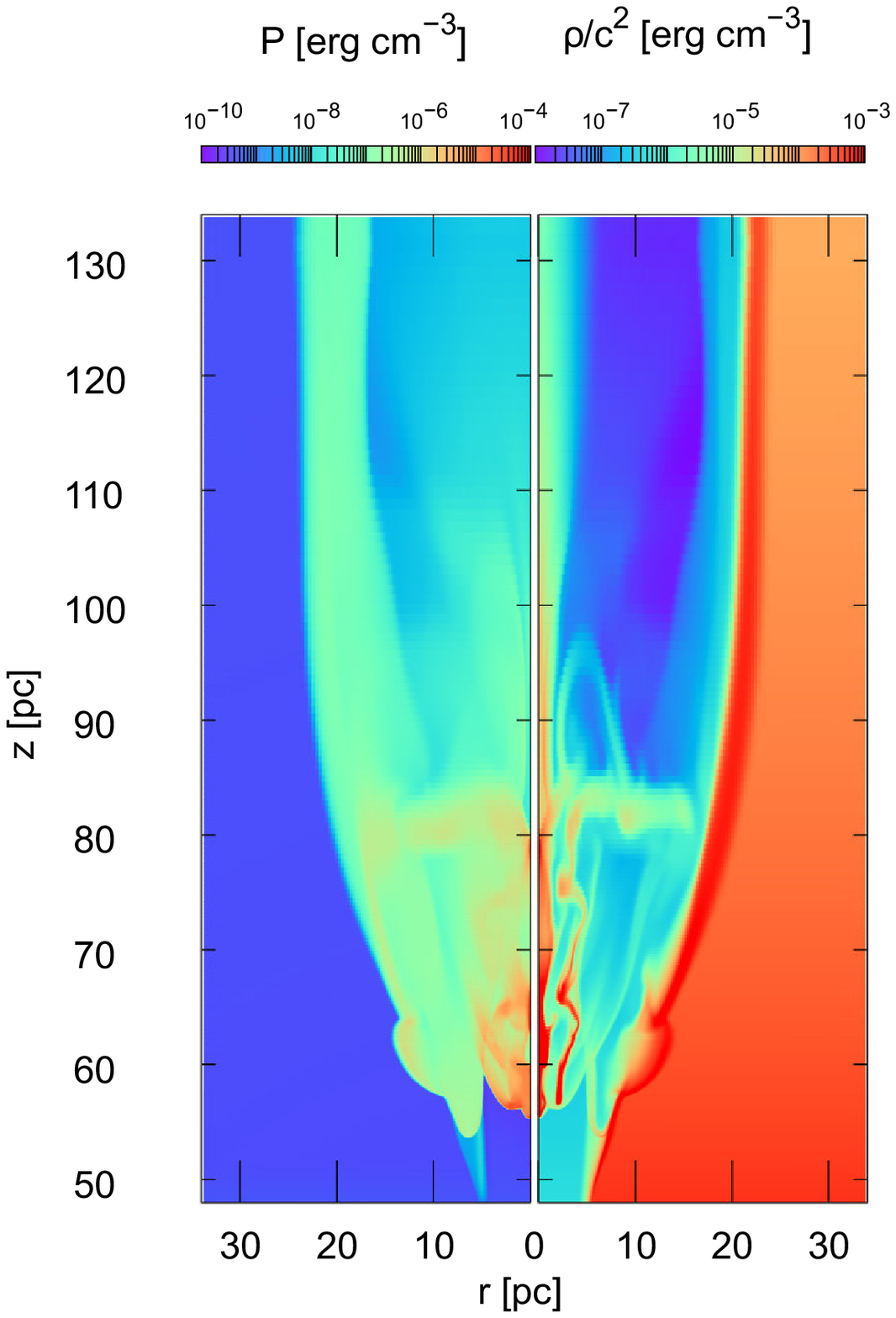} \hspace{20pt} 
\includegraphics[width=0.4\textwidth,keepaspectratio]{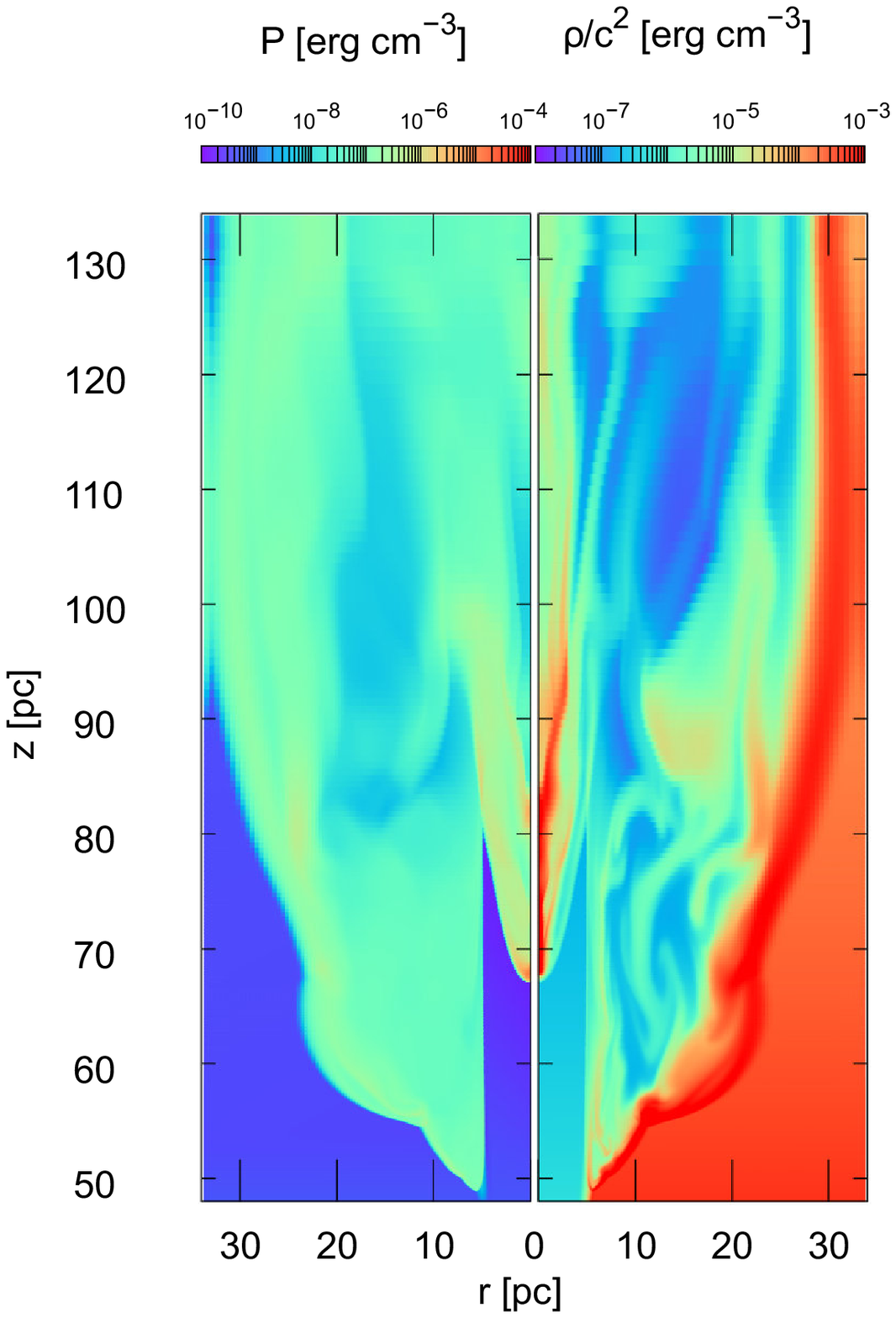} \hspace{20pt}
\caption{Combined colour maps of a SN material pressure (left) and density (right) at different times: beginning of the interaction (top left), $t=592.6$ yr (top right), $t=1188.1$ yr (bottom left), and $t=1848.9$ yr (bottom right).}
\label{fig:tracG}
\end{figure*}

To quantify the precision of the semi-analytic model used to describe the evolution of a SN ejecta accelerated by the jet (see Sect \ref{dynamic}), we have performed axisymmetric, relativistic hydrodynamical (RHD) simulations in two dimensions of the interaction between a jet and a spherical cloud. We have considered a jet of negligible thermal pressure (1\% of the jet ram pressure), $L_{\rm j} =10^{45}$~erg~s$^{-1}$, and $\Gamma_{\rm j} = 10$, and a uniform cloud at rest of $10 M_{\odot}$, initial radius $R_{\rm c} = 1.25 \times 10^{18}$ cm ($12.5$ cells), and in pressure balance with the jet ram pressure (see Sect.~\ref{dynamic}). At the considered interaction location, the magnetic field was assumed to be dynamically negligible. The code that solved the RHD equations was the same as in \citet{delacita17}: third order in space \citep{mignone2005}; second order in time; and using the Marquina flux formula \citep{donat1996,donat1998}. The adiabatic index of the gas was fixed to $4/3$, corresponding to an ideal, monoatomic relativistic gas.

The grid adopted consisted of a uniform grid with 150 cells between $r_{0}^{\rm grid} = 0$ and $r_{\rm max}^{\rm grid} = 1.5 \times 10^{19}$ cm in the $r$-direction, and 300 cells between $z_{0} ^{\rm grid} = 1.48 \times 10^{20}$ and $z_{\rm max}^{\rm grid} = 1.78\times 10^{20}$~cm in the $z$-direction. An extended grid was added with 150 cells in the $r$-direction, from $r_{\rm max}^{\rm grid} = 1.5 \times 10^{19}$ and $r_{\rm max}^{\rm ,grid}\approx 10^{20}$~cm, and with 200 cells in the $z$-direction, from $z_{\rm max}^{\rm grid} = 1.78 \times 10^{20}$ and $z_{\rm max}^{\rm ,grid}\approx 4.2\times 10^{20}$~cm. The resolution was chosen such that no significant differences could be seen in the hydrodynamical results when increasing the resolution.

Inflow conditions (the jet) were imposed at the bottom of the grid, reflection at the axis, and outflow in the remaining grid boundaries. On the scales of the grid, for simplicity we approximated the jet streamlines as radial, and added a smooth but thin shear layer transiting from the jet properties to the external medium properties (radial velocity of $10^8$~cm~s$^{-1}$, number density $\approx 1$~cm$^{-3}$, pressure equal to the jet thermal pressure) at $\theta\approx 1/\Gamma_{\rm j}$.
  
Figure \ref{fig:tracG} shows combined maps of pressure (left) and density (right) at different times, showing the beginning of the interaction (top left), and three intermediate stages:  $t=592.6$ yr (top right), $t=1188.1$ yr (bottom left), and $t=1848.9$ yr (bottom right). 
The plots show some of the effects discussed in Sects.~\ref{dynamic} and \ref{appL}, namely:

\begin{enumerate}
\item The SN ejecta completely covers the jet cross section from an early time;
\item Despite disruption, the cloud evolves roughly as a coherent structure (see also the figures shown below) until it has moved significantly further downstream;
\item The cloud does not expand much beyond the jet original radius before its disruption;
\item The jet begins to accelerate the SN material after $\sim 1000$ yr (similar to what is shown in Fig.~\ref{fig:Gc} for the jet with $L_{\rm j} =10^{45}$~erg~s$^{-1}$), although in fact the simulation acceleration time is a few times longer.
\end{enumerate}

\begin{figure*}[!ht]
\centering
\includegraphics[width=0.45\textwidth,keepaspectratio]{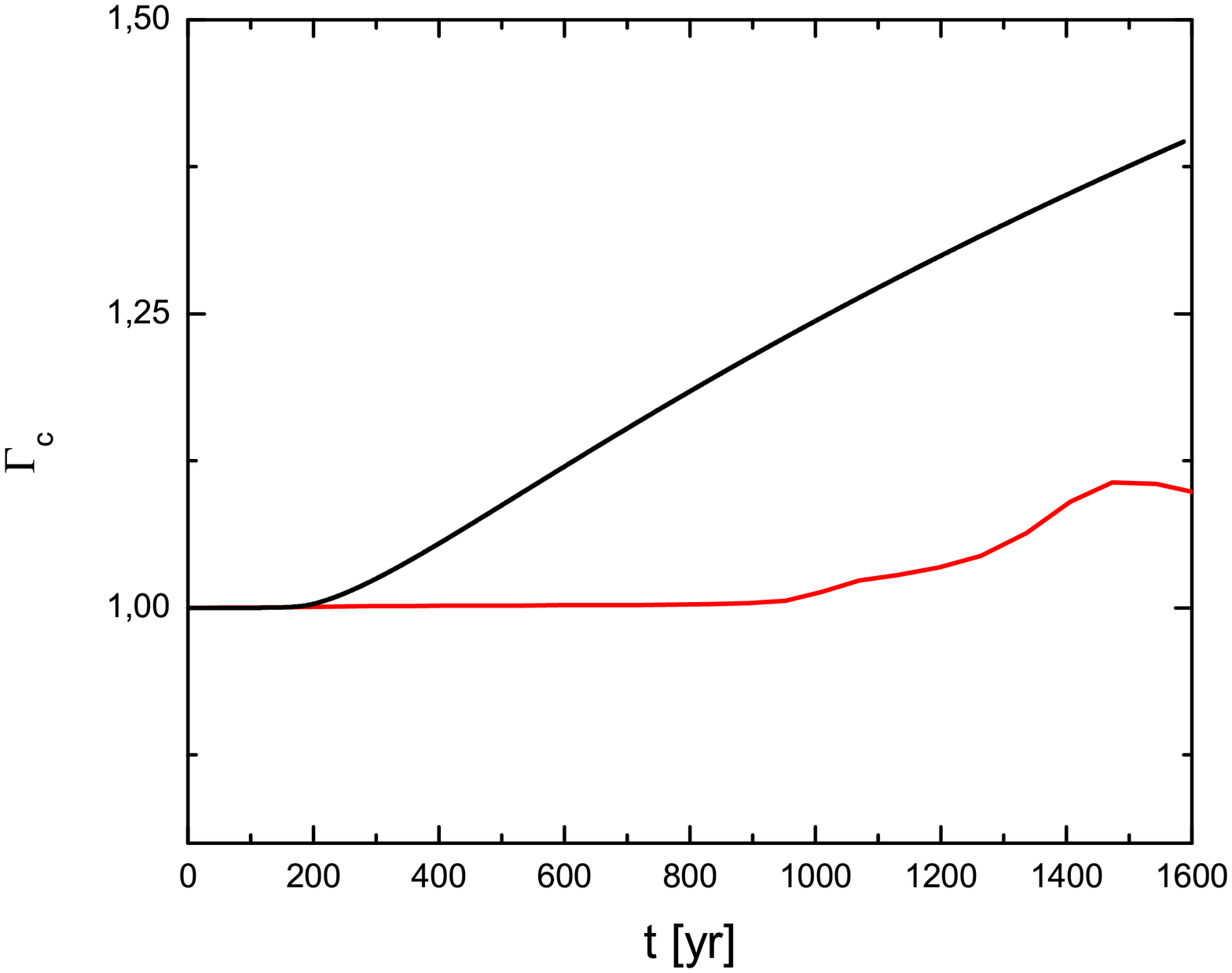}\hspace{20pt} 
\includegraphics[width=0.45\textwidth,keepaspectratio]{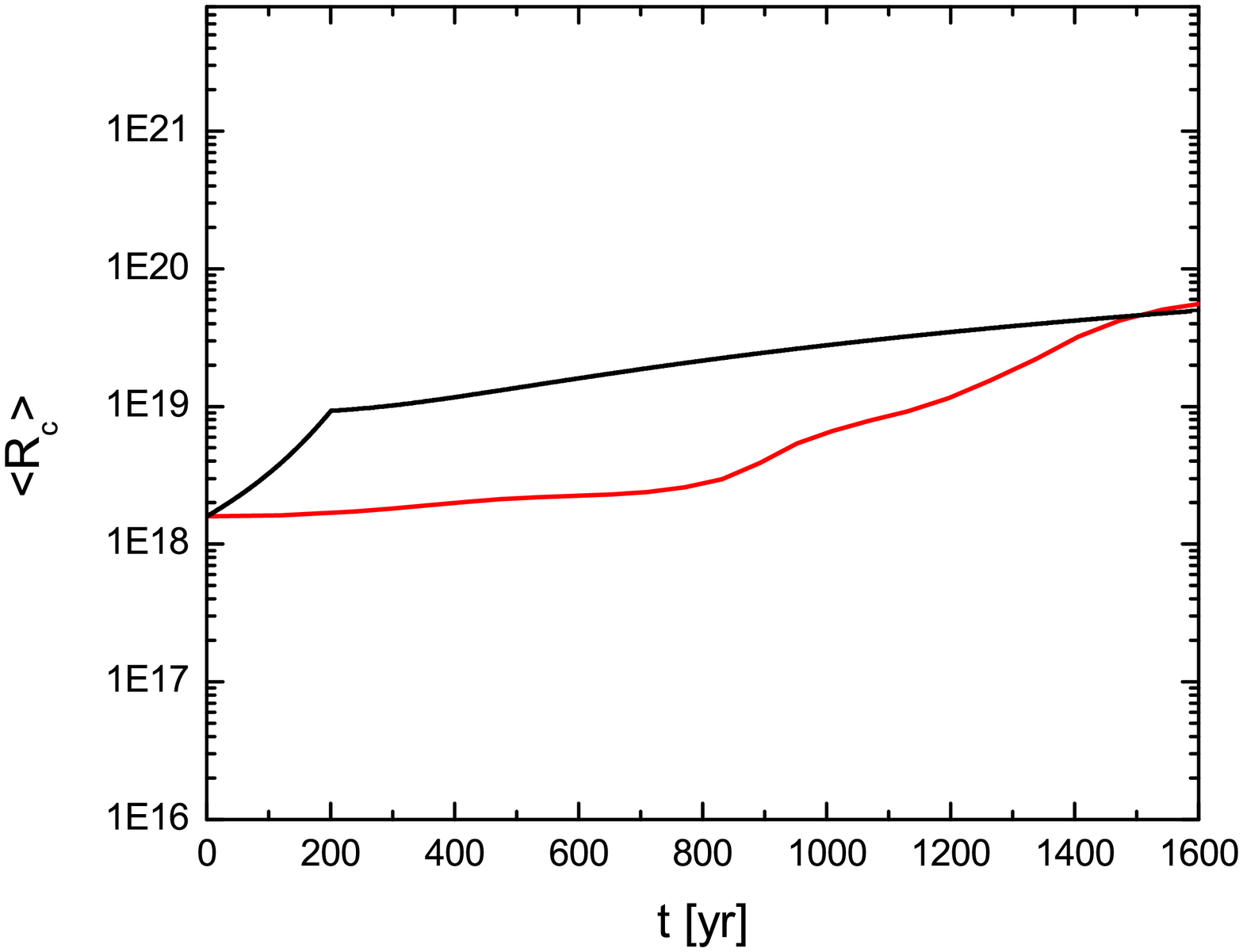}\hspace{20pt} 
\caption{Comparison between the cloud evolution obtained using the semi-analytical approach (black lines) and the average values evolution obtained in the numerical simulations (red lines) for the Lorentz factor (left), and mass-averaged cylindrical radius (right). The time shown in these plots is in the laboratory frame.}
\label{fig:comparacion}
\end{figure*}

To better illustrate the similarities between the semi-analytical treatment and the numerical simulations, in Fig. \ref{fig:comparacion} we show the comparison between the evolution of the main parameters of the cloud derived using both approaches. In the left panel we show the evolution of the cloud Lorentz factor; the acceleration of the cloud is reasonably reproduced by the treatment presented in Sect.~\ref{dynamic}, although, as mentioned above, the acceleration time obtained from the simulation is longer, favoring detectability (the duty cycle discussed in Sect.~\ref{dc} is a conservative estimate if the acceleration time is longer). In the right panel, we have plotted the evolution of the mass-averaged cylindrical radius of the cloud. This parameter differs somewhat from the spherical semi-analytical case, but the differences are small in the long run, and with the jet still being effectively fully covered by the SN ejecta, which means that our approximation should be accurate enough at this stage. A generalization of the simulation to include other cases, and the computation of the radiative outcome, are left for future work.

\end{appendix}

\end{document}